\documentclass[prd,showpacs,nofootinbib,preprint]{revtex4-1}
\usepackage{amsmath}
\usepackage{amsfonts}
\usepackage{autonum}
\usepackage{graphicx,xcolor}

\def\be{\begin{equation}}
\def\ee{\end{equation}}
\def\ba{\begin{eqnarray}}
\def\ea{\end{eqnarray}}

\def\lb{\label}

\def\J{{\cal A}}
\def\non{nonumber}

\def\lb{\label}

\def\k{\nu}
\def\Z{w}
\def\ta{{\hat \tau}}

\def\rh{{\hat \rho}}

\def\c{{\hat c}^\mu}

\newcommand{\sgn}{\operatorname{sgn}}
\newcommand{\Ai}{\operatorname{Ai}}

\numberwithin{equation}{section}

\begin{document}

\title{ 
 Synchrotron radiation in odd dimensions
}

\author{D.V. Gal'tsov}
\email{galtsov@phys.msu.ru}
\affiliation{Faculty of Physics, Moscow State University, 119899, Moscow, Russia}
\author{M. Khlopunov}
\email{khlopunov.mi14@physics.msu.ru}
\affiliation{Faculty of Physics, Moscow State University, 119899, Moscow, Russia}

\begin{abstract}
In odd space-time dimensions, the retarded solution of the massless wave equation has support not only on the light cone, but also inside it. At the same time, a  free massless field should propagate at the speed of light. The apparent contradiction of these two features is resolved by the fact that the emitted part of the field in the wave zone depends on the history of motion up to the retarded moment of proper time.
It is shown that in the case of circular motion with ultrarelativistic velocity, the main contribution to the radiation amplitude is made by a small interval of proper time preceding the retarded time, and thus the tail term is effectively localized. We obtain a tentative formula for scalar synchrotron radiation in $D$ dimensions: $P =g^2(\omega_0\gamma^2/\sqrt{3})^{D-2}$, which is  explicitly verified in $D=3,\,4,\,5$.  
\end{abstract}
 
\pacs{04.20.Jb, 04.50.+h, 04.65.+e}
\maketitle

\section{Introduction}

Over the past two decades, interest has arisen in the theory of radiation in space-time dimensions other than four. This was due to the emergence of theories with large extra dimensions~\cite{Rubakov}, the development of a holographic approach to the description of quark-gluon plasma~\cite{Cardoso:2013vpa,Arefeva2014},  the beginning of gravitational-wave astronomy and other reasons. The case of three-dimensional space-time  became relevant in connection with the development of field-theoretical models in condensed matter physics, such as  quantum Hall effect, high-temperature superconductivity, and graphene. Radiation is the main classical and quantum process of the interaction of charged particles with massless fields, which are basic ingredients of a number of theories. Therefore, it seems to be of primary importance to study the general laws of radiation in spacetimes  other than four-dimensional.

More than a hundred years ago, Paul Ehrenfest~\cite{Ehr} called the features of radiation in four-dimensional electrodynamics a unique manifestation of the dimensionality of space-time, so different are the physical effects and the theoretical description of radiation in other dimensions. General features of wave propagation in arbitrary dimensions were discussed in the classical collections of mathematical physics, such as Courant and Hilbert \cite{courant2008methods}, lectures of Hadamard~\cite{hadamard2014lectures}, the book by Ivanenko and Sokolov \cite{IvSo48}. It was found that there is a fundamental difference between even and odd dimensions consisting in the failure of the Huygens principle in the odd case. In  an odd-dimensional space-times, the signal from an instantaneous flash of current reaches an observer after an interval of time required for propagation of the signal at the speed of light, but then the tail is observed endlessly. In even-dimensional spaces, this is not so; an instant signal ends instantly at the observation point. The mathematical reason is that the retarded Green's function of the D'Alembert equation in odd dimensions has support localized not only on the light cone, but also inside it. 

There is an intriguing possibility of detecting an additional dimension in the five-dimensional theory of Randal-Sundrum and other similar theories  by scanning them with light signals~\cite{Barvinsky:2003jf}. The idea of direct experimental search for extra dimensions was  further discussed in anticipation of future development of LISA project~\cite{Deffayet:2007kf}. Propagation of gravitational waves in presence of extra dimension and possibility of observational effects was discussed more recently in \cite{Andriot:2017oaz}. Birth of the gravitational wave  astronomy gave a new impulse to search  for possibility to explore dimensionality of spacetime. According to some  of the theories with extra dimensions, gravitational waves simply would leak into them, causing  their weakening under propagation in the universe. This effect, however, was not found in  the GW170817 neutron stars coalescence event \cite{Chakravarti:2019aup}. Meanwhile, further investigations are active, see recent papers and reference therein ~\cite{Yu:2019jlb,Cardoso:2019vof,Kwon:2019gsa}. Relation of the extra dimensions to the black hole shadows is also worth to be mentioned~\cite{Vagnozzi:2019apd}.

Motivation for studying radiation in various dimensions comes also from holography, namely, the holographic modeling of quark-gluon plasma~\cite{Arefeva2014}. From the gravity side it involves  consideration of gravitational radiation in collisions of ultrarelativistic particles or approximating them shock gravitational waves on the background of anti-de Sitter~\cite{Cardoso:2013vpa}. Radiation in higher-dimensions is  relevant to the problem of creation of black holes in particle collisions in theories with extra dimensions~\cite{Galtsov:2010vtu,Berti:2010gx,Galtsov:2012pcw}. Note also an interesting interplay of even and odd-dimensional features encountered in the black hole-brane \cite{Frolov:2003mc} and the particle-brane  \cite{Galtsov:2015yyr,Galtsov:2017udh} systems.

As was said, behavior of massless fields and radiation processes in even and odd dimensions are rather different,
being more familiar in the even case. This is why  in the most of the existing literature only the even-dimensional radiation problems were considered~\cite{Kosyakov:1992,Kosyakov1999,Cardoso:2002pa,Gurses:2003cc,Cardoso:2007uy,Berti:2010gx} .
The case of odd dimensions was discussed mainly in the context of the radiation reaction problem \cite{Galtsov:2001iv,Kazinski:2002mp,Yar7,Yaremko:2007zz,Shuryak:2011tt,Yar12, Dai:2013cwa}, see also the reviews \cite{Kosyakov:51252,Kosyakov:2018wek}. Meanwhile, the radiation reaction in dimensions other than four, creates additional problems.  In the paper~\cite{Galtsov:2001iv}, devoted to radiation reaction in both the even and odd dimensions, it was concluded that the four is the unique dimension where the {\em renormalizable and local} Lorentz-Dirac-type equation~\cite{Dirac} exists, which properly accounts for  radiation friction. In higher even dimensions, the renormalization of the particle mass is insufficient; it is necessary to introduce higher derivative  counterterms in the Lagrangian to subtract all divergences. Although formally all needed subtractions  can be elegantly performed \cite{Kazinski:2002mp,Kazinski:2005gx} using the axiomatic of distributions \cite{Shilov}, new counterterms correspond to a bare “rigid particle” whose equations of motion contain higher derivatives. Thus, an initial theory without higher derivatives is not classically renormalizable.   In the lowest odd-dimensional case $2+1$, the renormalization of mass is enough, but, due to tail, the resulting equation is non-local (integro-differential). Thus, a fully consistent treatment of radiation reaction is only possible in four dimensions, in agreement with Ehrenfest's reasoning. 

Four-dimensional equations with tail terms describing radiation reaction in the curved space theory are quite familiar since DeWitt and Brehme's calculation~\cite{DeWitt:1960fc}. Moreover, this  became nowadays a fashionable theory due to its importance for gravitational wave astronomy, for a recent review see~\cite{Barack:2018yvs}. The tail terms in the curved spacetime are due to scattering of waves on the spacetime curvature. For this reason they are usually computed using perturbation theory. Contrary to this, tail terms in odd dimensions are analytically known in the closed form and they can be understood in terms of dimensional reduction/oxidation (see below). Note that in higher even dimensional curved spacetimes additional renormalizations and additional finite geometrical terms occur~\cite{Galtsov:2007zz}.

As far as the genuine radiation processes are concerned, with exception for the high-energy bremsstrahlung problem  insensitive to whether the number of dimensions is even or odd~\cite{Galtsov:2010vtu,Galtsov:2012pcw}, and a brief note~\cite{Spirin}, where the use of fractional derivatives for odd-dimensional Green's functions was proposed, to our knowledge, this paper is the first one  attempting to calculate radiation in odd dimensions using the standard wave zone approach \cite{LL}  including an explicit calculation of synchrotron radiation in three and five dimensions. We use Rohrlich's approach to radiation \cite{Rohrlich1961,rohr} and its refinement due to Teitelboim \cite{Teit} (see also~\cite{Kosyakov1999,Galtsov:2004uqu,Galtsov:2010tny}) to prove, that the  long-range component of the retarded potentials corresponds to radiation in the same way as it does in four dimensions. This component, however, depends on the history of the particle motion preceding the retarded time, but not at the retarded moment only, as it does in the even-dimensional spacetimes.   Our results were doubly checked by calculation of the spectral distributon of radiation, whose evaluation is insensitive to whether the dimension is even or odd. We expect they can be tested also within the quantum approach developed in~\cite{Birnholtz:2013ffa,Birnholtz:2015hua,Porto:2016pyg,Harte:2018iim}.
 
The paper is organized as follows. In section II we briefly recall Ivanenko and Sokolov's derivation of the recurrent relations between odd-dimensional Green's functions of D'Alembert equation and show that there are no divergences in the scalar field in the limit of static charge. Section III is devoted to the calculation of scalar synchrotron radiation in $(2+1)$-dimensional spacetime using the coordinate representation for the retarded Green's function in the wave zone and the Fourier spectral decomposition. In section IV we perform similar calculations in the $(4+1)$-dimensional spacetime. We find a universal formula for synchrotron radiation verified in $D=3,\,4,\,5$ and tentatively valid in any $D$. In the last section we  briefly formulate our results and discuss relation to other work.
\section{The setup}
We write the action of the massive relativistic particle interacting with the massless scalar field in $D=n+1$-dimensional Minkowski spacetime as 
\begin{equation}
\label{eq:s-p_action}
S= -  \int \,  (m+g \varphi(z))\,\sqrt{ \eta_{\alpha \beta} \dot{z}^{\alpha} \dot{z}^{\beta}} d \tau +\frac{1}{2\Omega} \int  \eta^{\mu \nu} \partial_{\mu} \varphi(x) \partial_{\nu} \varphi(x)\,d^{n+1} x  ,
\end{equation}
where $m$ is the particle's mass, $g$ is the scalar charge, $z^{\mu}(\tau)$ is the particle's worldline, $ \dot{z}^{\mu} (\tau) = d z^{\mu} (\tau) / d \tau$. Here $\Omega$ is the area of the $(n-1)$-dimensional sphere of unit radius:
\begin{equation}
\label{eq:sphere_area}
\Omega=\frac{2 \pi^{n/2}}{\Gamma(n/2)}.
\end{equation}
The Minkowski metric is $\eta_{\mu\nu}={\rm diag}(1,-1,\ldots ,-1)$.
Our choice of the scalar coupling constant $g$ differs from a more frequent definition $f=g/m$  (see, e.g., \cite{Breuer:1974uc}) in a way to make it non-zero in the massless limit $m\to 0$.

This action leads to the following wave equation for the scalar field
\begin{eqnarray}
\label{eq:sc_eq_motion}
\square \varphi(x) = - \Omega j(x), \\
\label{eq:sc_current}
j(x) = g \int d \tau \left( \dot{z}^{\alpha} \dot{z}_{\alpha} \right)^{1/2} \delta^{n+1} (x-z(\tau)),
\end{eqnarray}
where $j(x)$ is the current   and $\square=\partial^{\mu} \partial_{\mu}$ is D'Alembert operator. The canonical energy-momentum tensor of the   scalar field  is
\begin{equation}
\label{eq:EMT_sc_gen}
T_{\mu\nu} (x) = \frac{1}{\Omega} \left( \partial_{\mu} \varphi \partial_{\nu} \varphi - \frac{1}{2} g_{\mu\nu} \partial_{\alpha} \varphi \partial^{\alpha} \varphi \right).
\end{equation}
\subsection{Green functions  in  odd dimensions}
For reader's conveniencs, we recall here Ivanenko and Sokolov's derivation of the recurrent relation for odd-dimensional Green functions of the scalar D'Alembert equation.
We present the retarded solution of the Eq. \eqref{eq:sc_eq_motion} as
\begin{eqnarray}
\label{eq:sq_eq_gen_sol}
\varphi (x) = - \Omega \int d^{n+1} x' G_{ret}^{n+1} (x-x') j(x'), \\
\label{eq:gr_fn_eq}
\square G_{\rm ret}^{n+1} (x-x') = \delta^{n+1} (x-x'),
\end{eqnarray}
where the retarded Green's function of the D'Alembert  equation is
\begin{equation}
\label{eq:ret_gr_fn_fourier}
G_{\rm ret}^{n+1} (x-x') = - \int \frac{d^{n+1} k}{(2\pi)^{n+1}} \frac{e^{-ik(x-x')}}{k^2+i\varepsilon k^0},
\end{equation}
with $k^2=k_\mu k^\mu$ and   $\varepsilon = +0$ defining the correct shift of a pole in the complex plane $k^0$. After integration over $k^0$, we obtain an integral over the Euclidean $n$-dimensional space:
\begin{eqnarray}
\label{eq:sp_int_gr_fn}
G_{\rm ret}^{n+1} (x-x') = \int \frac{d^n k}{(2\pi)^n}  \frac{\sin \omega T}{ \omega }\;{\rm e}^{i \mathbf{k} \mathbf{R} }, \\
\label{eq:vect_not}
\mathbf{R} = \mathbf{r} - \mathbf{r}'; \; T=x^0 - {x'^0}; \; \omega = |\mathbf{k}|.
\end{eqnarray}
One can introduce the hyperspherical coordinates \cite{2009arXiv0901.3488L} and integrate over the cyclic angles, obtaining
\begin{equation}
G_{\rm ret}^{n+1} (x-x') = \frac{2 \pi^{(n-1)/2}}{\Gamma((n-1)/2)} \int \frac{d \omega d \theta_{n-2}}{(2\pi)^n} \omega^{n-2} \sin (\omega T) \sin^{n-2} (\theta_{n-2}) e^{i\omega R \cos \theta_{n-2}}.
\end{equation}
The remaining angular integral can be expressed through the Bessel function of the order $\nu-1$, with $n=2\nu$ \cite{watson,zwillinger2014table}:
\begin{equation}
G_{\rm ret}^{2\nu+1} (x-x') = \frac{R}{(2\pi R)^{\nu}} \int \limits_{0}^{\infty} d \omega \, \omega^{\nu-1} J_{\nu-1} (\omega R) \sin \omega T.
\end{equation}

Using the recurrent relations between the Bessel functions
\begin{equation}
\left( \frac{d}{x dx} \right)^{m} \frac{J_n (x)}{x^n} = (-1)^m \frac{J_{n+m} (x)}{x^{n+m}},
\end{equation}
we obtain the following  generating formula:  
\begin{equation}
\label{eq:odd_gen_gr_fn}
G_{\rm ret}^{2\nu+1} (X) = \frac{(-1)^{\nu-1}}{(2\pi)^{\nu-1}} \frac{d^{\nu-1}}{(RdR)^{\nu-1}} G_{\rm ret}^{2+1} (X).
\end{equation}
For $\nu=1$ one has
\begin{equation}
\label{eq:polar_rep_gr_fn}
G_{\rm ret}^{2+1} (x-x') = \frac{1}{2 \pi} \int \limits_{0}^{\infty} d \omega \, J_{0} (\omega R) \sin \omega T.
\end{equation}
This integral gives the Heaviside function \cite{IvSo48,watson}, so we obtain:
\begin{equation}
\label{eq:2+1_gr_fn}
G_{\rm ret}^{2+1} (X) = \frac{\theta(X^0)}{2\pi} \frac{\theta (X^2)}{\sqrt{X^2}},
\end{equation}
where  $X^{\mu}=x^{\mu}-x'^{\mu}$. This  function is localised on and inside the future light-cone $X^2=0$. Note that the denominator vanishes  there, but, as we will see later, this does not imply divergence of the retarded solution.

From the recurrent formula it is clear that  in any odd dimensions the retarded Green function has support on and inside the future light cone. Performing differentiation, we obtain the sum of terms each of which has zeroes in the denominator at $X^2=0$. However the full Green function should not have infinities other than delta-function, so cancellation of divergences between different terms in this sum is expected.  
\subsection{Cancellation of divergences in the static limit}
Let us verify the absence of divergences in the field of a static particle in the case of $(4 + 1)$ dimensions, which is the most transparent illustration.

From $\eqref{eq:odd_gen_gr_fn}$ we find   the retarded $(4+1)$-dimensional   Green  function as a sum of two terms 
\begin{equation}
\label{eq:5d_gr_fn}
G_{\rm ret}^{4+1} (X) = \frac{\theta (X^0)}{2 \pi^2} \left( \frac{\delta (X^2)}{(X^2)^{1/2}} - \frac{1}{2} \frac{\theta (X^2)}{(X^2)^{3/2}} \right),
\end{equation}
each of which has zero in the denominator on the light cone $X^2=0$.
We will use the "finite lifetime" trick to demonstrate   cancellation of divergences between  two terms in the static limit. For this,  first assume that the source is switched on for a finite interval of the proper time $\tau \in \lbrack a,b \rbrack$, with  $a<0$ and $b>0$. On this interval one has $z^\mu (\tau)= [ \tau,0,0,0,0 ] $, so from the Eqs. $\eqref{eq:sc_current}$ and $\eqref{eq:sq_eq_gen_sol}$ with account for $\eqref{eq:sphere_area}$ we obtain:
\begin{equation}
\label{eq:gen_fl_ret_f}
\varphi  (x) = \frac{g}{2} \int \limits_{a}^{b} d \tau \left( \frac{\theta (t-\tau-r-\varepsilon)}{\lbrack (t-\tau)^2 - r^2 \rbrack^{3/2}} - \frac{\delta (t-\tau-r-\varepsilon)}{r \lbrack (t-\tau)^2 - r^2 \rbrack^{1/2}} \right),
\end{equation}
where we introduced a regularizing parameter $\varepsilon>0$, shifting the singularities from the light-cone. Performing an integration, one finds:
\begin{equation}
\label{eq:fl_ret_f}
\varphi  (t,r) = \frac{g}{2} \begin{cases} 0, \quad t < a+r, \\
\displaystyle -  \frac{(t-a)}{r^2 \lbrack (t-a)^2 - r^2 \rbrack^{1/2}}, \quad t \in [ a+r ,\, b+r ), \\
\displaystyle  \frac{(t-b)}{r^2 \lbrack (t-b)^2 - r^2 \rbrack^{1/2}} - \frac{(t-a)}{r^2 \lbrack (t-a)^2 - r^2 \rbrack^{1/2}} , \quad t \geq b+r.
\end{cases}
\end{equation}
Passing   to the  limit  of an eternal particle worldline $a \to - \infty, \, b \to \infty$,  one gets at any $t$ a finite result
\begin{equation}
\varphi  = - \frac{g}{2r^2}.
\end{equation}

Similarly, divergences in the sum representation of higher $\nu$ odd-dimensional Green functions are expected to mutually cancel and not only in the static case. We will show this explicitly for $D=3,\,5$.   
\subsection{Retarded field in the wave zone}

Recall that in four dimensions the retarded electromagnetic field of a point charge consists of two parts: one proportional to $1/r^2$ and representing the deformed Coulomb field, and another, acceleration dependent, which falls down as $1/r$. The second gives a non-zero flux of the field energy-momentum (Pointing vector) through the distant sphere, and thus represents radiation. To argue that this is radiation indeed, 
Rohrlich \cite{Rohrlich1961,rohr}   and Teitelboim \cite{Teit}  (see also \cite{Kosyakov1999, Kosyakov:51252,Galtsov:2004uqu,Kosyakov:2018wek}), computed the most long-range part of the on shell energy-momentum tensor, showing that it  exhibits special properties, meaning that the corresponding part of the field energy-momentum propagates at the speed of light. 
Similar decomposition and reasoning holds for the gradient of the retarded scalar field of the scalar charge. All this remains valid in any spacetime dimensions, with the difference that in $D$ dimensions the area of the far sphere grows with distance as $r^{D-2}$, so the relevant asymptotic behavior of the field gradient in the wave zone is $1/r^{D/2-1}$. Note that in odd dimenions this power is half-integer. 

In the Rohrlich-Teitelboim construction the use of certain covariantly defined quantities seems essential, so we briefly recall their definition.
Consider a pointlike scalar charge moving along a worldline $z^{\mu}(\tau)$ with the $D$-velocity $v^{\mu}=dz^{\mu}/d\tau$, and denote the coordinates of the observation point as $x^{\mu}$. Consider the observation point as a top of the light cone in the past, and denote the intersection point of  the light cone with the world line of a particle as
$\hat{z}^{\mu}={z}^{\mu}(\hat{\tau})$, where $\hat{\tau}$ is the moment of proper time corresponding to the emission of a signal propagating  to the observation point at the speed of light. The quantity $\hat{\tau}$ is called the retarded proper time; it is determined by the equation
\begin{equation}
\label{eq:ret_prop_time_eq}
( x^{\mu} - z^{\mu} (\hat{\tau}) )^2 = 0.
\end{equation}
In what follows, all hatted quantities will correspond to the  retarded proper time $\hat{\tau}$. We then introduce two spacetime vectors: a lightlike vector $\hat{R}^{\mu}=x^{\mu}-\hat{z}^{\mu}$ directed from $\hat{z}^{\mu}$ to the observation point, and a spacelike unit vector $\hat{u}^{\mu}$, orthogonal to $\hat{v}^{\mu}$. The sum of the vectors $\hat{v}^{\mu}$ and $\hat{u}^{\mu}$ forms a lightlike vector $\hat{c}^{\mu}=\hat{v}^{\mu}+\hat{u}^{\mu}$. These vectors have the following properties:
\begin{align}
\label{eq:ret_q_1}
&\hat{v}^2 = - \hat{u}^2 = 1; \quad \hat{c}^2 = 0; \quad \hat{c} \hat{v} = - \hat{c} \hat{u} = 1; \quad \hat{v} \hat{u} = 0, \\
\label{eq:ret_q_2}
&\hat{R}^{\mu} = \hat{\rho} \hat{c}^{\mu}; \quad \hat{\rho} = \hat{v} \hat{R}; \quad \hat{R}^2 = 0.
\end{align}
It is worth noting that $\hat{\rho}$, being the scalar product of two spacetime vectors, is a Lorentz-invariant distance, equal to the distance in the Lorentz frame comoving with the charge at the retarded moment $\ta$. Note also that for a point charge, moving along the world-line $z^\mu(\tau)$ for an infinite proper time $-\infty<\tau<\infty$, certain care is needed to correctly define the asymptotic conditions for acceleration, for details see \cite{Teit}. Here we will not discuss this subtlety, considering the simple case of  periodic motion along a circle. Far from the circle,  $\rh\sim r$, so the Lorentz-invariant definition of the distance is equivalent to the naive definition. But in order to obtain Lorentz covariant expansions of tensors, one has to use $1/\rh $ as an expansion parameter in even dimensions and $1/\rh^{1/2}$ in odd. 
 
Now we come back to an asymptotic structure of the on-shell energy-momentum tensor computed with the retarded solutions of the wave equation.
Recall that in four-dimensional electrodynamics Teitelboim \cite{Teit} demonstrated that the following decomposition holds (here we use slightly different notation):
\be
T^{\mu\nu}=T_{\rm Coul}^{\mu\nu}+ T_{\rm mix}^{\mu\nu}+ T_{\rm rad}^{\mu\nu},
\ee
where the Coulomb part falls down at spatial infinity as $\rh^{-4}$, the mixed part -- as $\rh^{-3}$, and the last part as $\rh^{-2}$. 
Teitelboim found that in four dimensions the most long-range term of the on-shell energy-momentum tensor has the following properties:
\begin{itemize} 
\item 
It is separately conserved $\partial_\nu T_{\rm rad}^{\mu\nu}=0$.
\item
It is proportional to the direct product of two null vectors $c^\mu c^\nu$, and therefore $c_\mu T_{ \rm rad}^{\mu\nu}=0$.
\item 
It falls down  as $1/\hat{\rho}^2$ and gives positive definite flux through the distant sphere.
\end{itemize}
It is clear that this tensor  corresponds to propagation of the field energy-momentum with the speed of light. Thus, the radiation power can be computed as the flux of the energy, associated with $T_{\rm rad}^{\mu\nu}$. Similar structure holds in the scalar theory.

Now we pass to our theory of the scalar radiation in $D$ dimensions. We substitute the retarded solution of the wave equation into the bilinear functional (\ref{eq:EMT_sc_gen}) and expand this quantity in inverse powers of $\rh$. The result is as follows. The most short-range term decays as $\rh^{4-2D}$ in all dimensions, the mixed part is absent for $D=3$ and consists of more than one term for $D>4$, varying from $\rh^{5-2D}$ to $\rh^{1-D}$,  the most long-range part decays as $\rh^{2-D}$, as expected for radiation. All the listed properties of the last term $T_{ \rm rad}^{\mu\nu}$ of this expansion hold in any $D$.

This remains true both in even and odd dimensions.  An essential difference, however, is that  the radiated field in the  even-dimensional case depends on the particle kinematic quantities (velocity, acceleration and possibly higher derivatives of the velocity) at the retarded moment of the proper time $\ta$ only, while in the odd case it  depends on the entire history of motion before and at this moment.
Nevertheless, due to the above properties, the  energy-momentum associated with the long range component of the energy-momentum tensor will propagate at the speed of light.

It is  worth noting once again that the retarded field, in accordance with the explicit form of the Green functions (\ref{eq:odd_gen_gr_fn}), will have at large $\rh$ an expansion in terms of the half-integer powers   $1/\rh^{1/2}$ in odd dimensions.

The flux of the radiated energy-momentum passing per unit time through the  $(2\k-1)$-dimensional sphere of radius $r$ will be  given by the integral
\be\lb{W}
W_{2\k+1}^\mu = \int \, T_{\rm rad}^{\mu i}\; n^{i}\, r^{2\k-1} \, d\Omega_{2\k-1}, \quad i=\overline{1,2\k},
\ee
where $d\Omega_{2\k-1}$ is an angular element, and $\bf{n}$ is a unit spacelike vector in the direction of observation.
 
Using this approach, we calculate scalar synchrotron radiation from a circularly moving particle in $2 + 1$ and $4 + 1$ dimensions. To verify the correctness of this calculation, we also compute the total radiation power using Fourier
spectral decomposition, bearing in mind that this second method of  does not depend on
whether the dimension of spacetime is even or odd.

\subsection{Spectral decomposition}
Spectral representation can be introduced in a universal way in any dimensions, both  even and odd. In terms of the Fourier transforms defined as
\begin{eqnarray}
\label{eq:sc_fourier}
\varphi (x) = \int \frac{d^{n+1}k}{(2\pi)^{n+1}} e^{-ikx} \tilde{\varphi} (k), \\
\label{eq:sc_current_fourier}
j (x) = \int \frac{d^{n+1}k}{(2\pi)^{n+1}} e^{-ikx} \tilde{j} (k),
\end{eqnarray}
where $kx=k_\mu x^\mu$ with the $n+1$-dimensional wave-vector $k^\mu$,  the retarded/advanced solutions of the D'Alembert equation read
\begin{equation}
\label{eq:ret_sc_fourier}
\tilde{\varphi}_{\rm \pm} (k) = \frac{\Omega \tilde{j} (k)}{k^2 \pm i \varepsilon k^0} = \Omega \tilde{j} (k) \left \lbrack   \frac{{\cal P}}{k^2} \mp i \pi \sgn (k^0) \delta (k^2) \right \rbrack.
\end{equation}
The total loss of the energy-momentum due to radiation can be presented as the work done by the current in the field presented by the half-difference of the retarded and advanced solutions of the wave equation \cite{Dirac}:
\begin{equation}
\label{eq:rad_en_mom_fourier}
P_{\mu} = \frac{i}{2} \int \frac{d^{n+1}k}{(2\pi)^{n+1}} \left[\tilde{\varphi}_+ (k)- \tilde{\varphi}_- (k)\right]\tilde{j} (-k) k_{\mu}=
 \frac{\Omega}{(2\pi)^n} \int d^{n+1}k \, k_{\mu} |\tilde{j}(k)|^2 \theta (k^0) \delta (k^2).
\end{equation}
Inserting the Fourier transform of the scalar current into the Eq. $\eqref{eq:rad_en_mom_fourier}$ and passing to integration over $t=\tau\gamma$ instead of $\tau$ we find the following  representation for the spectral-angular distribution of the total energy radiated:  
\begin{equation}
\label{eq:gen_sp-ang_en_dist}
\frac{dP_0}{d\omega d\Omega} = \frac{\Omega \omega^{n-1} g^2}{2 (2\pi)^{n} \gamma^2} \left \vert \int \limits_{-\infty}^{+\infty} dt {\rm e}^{i (\omega t - \mathbf{k} \mathbf{z} (t))}  \right\vert^2,
\end{equation}
where $\omega=|\mathbf{k}|$. 
This generalizes the well-known $D=4$ formula \cite{Baier} to arbitrary dimensions. 
\section{$\bold{(2+1)}$ theory}
\subsection{Coordinate representation}
The retarded  solution of the Eq. $\eqref{eq:sc_eq_motion}$  is  given by $\eqref{eq:sq_eq_gen_sol}$:
\begin{eqnarray}
\label{eq:2+1_gen_sol}
\varphi^{\rm ret} (x) = - 2 \pi \int G_{\rm ret}^{2+1} (X) j(x') d^{2+1} x', \\
G_{\rm ret}^{2+1} (X) = \frac{\theta (X^0)}{2\pi} \frac{\theta (X^2)}{\sqrt{X^2}},
\end{eqnarray}
with the current (\ref{eq:sc_current}). Therefore we obtain:
\begin{equation}
\label{eq:2+1_sc_int}
\varphi^{\rm ret} (x) = - g \int d \tau \frac{\theta (X^0(z)) \theta (X^2(z))}{\sqrt{X^2(z)}},
\end{equation}
where the vector $X^{\mu}(z)=x^{\mu}-z^{\mu}(\tau)$ joins the observation point and an {\em instantaneous} position of a particle. In fact,  we need the gradient of the retarded field with respect to the observation point:
\begin{equation}
\label{eq:grad_int}
 \varphi_\mu^{\rm ret} \equiv
\frac{\partial \varphi^{ \rm ret}}{\partial x^\mu}= 2g \int d \tau\; \theta (X^0(z)) \left( \frac{1}{2} \frac{\theta (X^2(z))}{(X^2(z))^{3/2}} - \frac{\delta (X^2(z))}{(X^2(z))^{1/2}} \right) X_{\mu}(z).
\end{equation}
To see the asymptotic behavior of this quantity at large distances $r\gg R_0$, it is convenient to present $X^\mu$ as 
\be 
\label{eq:Z_def}
X^{\mu}  = x^{\mu} - z^{\mu}(\tau) = \Z^{\mu}  +{\rh}\c,\quad \Z^{\mu}=z^{\mu}(\ta)-z^{\mu}(\tau),
\ee
and then expand all quantities in terms of the small ratio $\Z^\mu/\rh$.
In the leading approximation,  $X^2\sim 2\Z_\mu \c=2(\Z\hat{c})$. Using also the relation  
\begin{equation}
\label{eq:d-t_rel}
\theta(X^0) \delta(X^2) = \frac{\delta (\tau - \hat{\tau} )}{2\hat{\rho}},
\end{equation}
we arrive at the following expression for the leading term $\varphi_{\mu}^{\rm rad}$ of the asymptotic expansion      of  $\varphi_{\mu}^{\rm ret}$ in $(1/\rh)^{1/2}$:
\begin{equation}
\label{eq:phi_gen_int}
\varphi_{\mu}^{\rm rad} = \frac{g\hat{c}_{\mu}}{2^{1/2}\hat{\rho}^{1/2}} \int \limits_{-\infty}^{\hat{\tau}} d \tau \left( \frac{1}{2 (\Z\hat{c})^{3/2}} - \frac{\delta(\tau - \hat{\tau})}{(\Z\hat{c})^{1/2}} \right).
\end{equation}
One can see that it is given by an integral over all the past worldline prior to the retarded time $\tau\leq \ta$. This is what happens in all odd dimensions: the radiation field in the wave zone is collected from  the entire history of motion preceeding  $\ta$. 
This is similar to the tail in the radiation amplitude in four-dimensional curved spacetime \cite{DeWitt:1960fc}, where it is due to scattering of radiation on the spacetime curvature. But, in the odd-dimensional flat spacetime, the origin of tail is different. As was explained in \cite{Galtsov:2001iv}, one can think of the $D$-theory as dimensionally reduced $D+1$-theory of paralles wires, whose projections are seen as point particles in $D$.  
Then, the radiation signal at  infinity of the $D$-world will be collected from all pieces of wires of the $D+1$-world, in which the propagation occurs at the speed of light, but the spatial distance from the given segment of the wire to an observation point is greater (or equal at the limiting point) than its projection onto $D$-world. Integration over the wires will produce a tail in the $D$ theory.
 
Each of two integrals in (\ref{eq:phi_gen_int}) diverges when $(\Z {\hat c})=0$. This happens in the limit $\tau\to\ta$, where $(\Z_\mu \c)\to (u_\mu \c)(\ta-\tau)=(\ta-\tau)$. To regularize the integral with the delta-function, we shift $\tau\to \tau+\varepsilon$ with $\varepsilon\to 0^+$, obtaining  
\begin{equation}
\label{eq:delta_transform}
\int \limits_{-\infty}^{\hat{\tau}} d \tau \frac{\delta (\tau - \hat{\tau} + \varepsilon)}{(\Z\hat{c})^{1/2}} = \frac{1}{\varepsilon^{1/2}},
\end{equation}
that can be rewritten as
\begin{equation}
\label{eq:2+1_st_contr}
\frac{1}{\varepsilon^{1/2}} = \frac{1}{2} \int \limits_{-\infty}^{\hat{\tau} - \varepsilon} \frac{d \tau}{(\hat{\tau} - \tau)^{3/2}}.
\end{equation}
Using this representation   of the delta-function term, we arrive at the finite quantity
\begin{equation}
\label{eq:2+1_rad_sc_gr}
\varphi_{\mu}^{\rm rad}  =\lim_{\varepsilon\to 0} \frac{g\hat{c}_{\mu}}{2^{3/2} \hat{\rho}^{1/2}} \int \limits_{-\infty}^{\hat{\tau} - \varepsilon} d \tau \left( \frac{1}{(\Z\hat{c})^{3/2}} - \frac{1}{(\hat{\tau}-\tau)^{3/2}}
\right),
\end{equation}
where, for brevity, we can omit the symbol $ \varepsilon $ at the top point of integration, remembering that we need to perform some transformation of the integrand (usually integrating by parts) to make its finiteness manifest.  It can be shown that, for the particle motion with constant velocity, the radiated part of the field gradient vanishes, as expected.
 
The long range energy-momentum tensor $T_{\mu\nu}^{\rm rad}$ can be found substituting the obtained expression into the bilinear form $\eqref{eq:EMT_sc_gen}$, leading to
\begin{equation}
\label{eq:EMT_2+1_gen}
T_{\mu\nu}^{\rm rad} = \frac{g^2 \hat{c}_{\mu} \hat{c}_{\nu} }{16 \pi \hat{\rho}} \J^{2} (x),
\end{equation}
where an integral radiation amplitude is introduced  
\begin{equation}
\label{eq:rad_int_start}
\J  = \int \limits_{-\infty}^{\hat{\tau}} \left( \frac{1}{(\Z\hat{c})^{3/2}} - \frac{1}{(\hat{\tau}-\tau)^{3/2}} \right),
\end{equation}
All physical information is contained in its first term, depending on the $z(\tau)$, while the second term just subtracts the divergence of the first on the upper limit. 
Clearly, the energy-momentum tensor obtained satisfies all the requirements of the Teitelboim definition in  the $(2+1)$-dimensional spacetime.
\subsection{Synchrotron radiation}
Now we proceed to calculate radiation from the circularly moving charge.
The particle's worldline $z^\mu(\tau)$ in terms of the proper time $\tau$  will read:
\begin{equation}
\label{eq:2+1_worldline}
z^{\mu} (\tau) = \left[\gamma \tau , R_{0} \cos ( \omega_{0} \gamma \tau) , R_{0} \sin (\omega_{0} \gamma \tau) \right],
\end{equation}
where $\gamma=E/m$ is the Lorentz factor of particle, $R_{0}$ is the radius of a circle  and $\omega_{0}$ is the  frequency of rotation. The corresponding  three-velocity is
\begin{equation}
\label{eq:2+1_velocity}
v^{\mu} (\tau) = \gamma\left[ 1, -  v \sin( \omega_{0} \gamma \tau) , v \cos (\omega_{0} \gamma \tau )\right],
\end{equation}
where $v=R_{0} \omega_{0}$, so that $\gamma=(1-v^2)^{-1/2}$. 

It will be convenient to express the retarded proper time $\hat{\tau}$ as a function of the coordinate time $t$ and the distance to the observation point from the {\em center} of the particle trajectory, which will be denoted as $R$. Using $\eqref{eq:ret_prop_time_eq}$  we find
\begin{equation}
\label{eq:ret_prop_time_exp}
\hat{\tau} = \frac{t-R}{\gamma}.
\end{equation}
Substituting this into the Eq. $\eqref{eq:ret_q_2}$, we arrive at the following expressions for  $\hat{\rho}$ and $\hat{c}^{\mu}$:
\begin{eqnarray}
\label{eq:2+1_rho}
\hat{\rho} = \gamma R \left( 1 + v \sin ( \omega_{0} \gamma \hat{\tau} - \phi ) \right), \\
\label{eq:2+1_c}
\hat{c}^{\mu} = \frac{R}{\hat{\rho}} [ 1 , \cos \phi , \sin \phi ],
\end{eqnarray}
where we have introduced the polar coordinates for the observation point: $x^{\mu} = [ t, \,R \cos \phi ,\, R \sin \phi ]$. Using the parameterization of the world-line $\eqref{eq:2+1_worldline}$ and introducing a new angular variable and   the integration parameter $$a=\omega_{0} \gamma \hat{\tau} - \phi + \pi/2,\quad s=\omega_{0} \gamma (\hat{\tau}-\tau),$$ we can present the contraction  $\Z_{\mu}\hat{c}^{\mu}$, entering  the  amplitude as follows:
\begin{equation}
\label{eq:2+1_Zc}
\Z\hat{c} = \frac{s - v \sin a - v \sin (s-a)}{\omega_{0} \gamma (1 - v \cos a)}.
\end{equation}
This gives the following  representation:
\begin{equation}
\label{eq:rad_int_circ}
\J  = (\omega_{0} \gamma)^{1/2} \int \limits_{0}^{+\infty} d s \left \lbrace \frac{ \left( 1 - v \cos a \right)^{3/2}}{\left( s - v \sin a - v \sin(s-a) \right)^{3/2}} - \frac{1}{s^{3/2}} \right \rbrace.
\end{equation}

\subsection{The ultrarelativistic case}
It can be expected that, as in the case of synchrotron radiation in four dimensions,  the radiation will be beamed and the integral amplitude can be simplified in the ultrarelativistic case $\gamma\gg1$.
Indeed, from the Eqs. $\eqref{eq:2+1_rho}$ and $\eqref{eq:rad_int_circ}$ with account for definitions of $s$ and $a$ we can see that the main part of radiation is then formed during a small interval of proper time before the retarded time $\hat{\tau}$ (in our new notation this corresponds to small $s$). One can also notice that in this case radiation is  beamed in the direction of the particle velocity  which corresponds to   $a=0$ or $\phi =\omega_{0} \gamma \hat{\tau} + \pi / 2$. 
This can be shown analysing an equation for $\hat{\tau}$, similarly to an analysis in
\cite{Shuryak:2011tt}. Namely, from the Eq. $\eqref{eq:ret_prop_time_eq}$ we  find the  relation
\begin{equation}
\gamma \frac{d \hat{\tau}}{d \phi} = \frac{R_{0} \sin (\omega_{0} \gamma \hat{\tau} - \phi)}{1+v \sin (\omega_{0} \gamma \hat{\tau} - \phi)},
\end{equation}
which in the new variables reads
\begin{equation}
\label{eq:phi-a_relation}
\frac{d a}{d \phi} = - \frac{1}{1-v \cos a}.
\end{equation}
This indicates that for  $v \to 1$ radiation is beamed within an angle $\delta \phi \sim \delta a / \gamma^2$. Expanding the denominator of the first term of the integrand in $\eqref{eq:rad_int_circ}$ in  Taylor series at $s=a=0$, we find that it has  minimum of width  $\delta s \sim \delta a \sim 1 / \gamma$. We can therefore find the amplitude in the leading-$\gamma$ approximation similarly to the case of the four-dimensional theory.
The leading  contribution to the radiation integral $\eqref{eq:rad_int_circ}$ will be 
\begin{align}
\label{eq:2+1_rad_int_lead}
&\J  = \gamma \omega_{0}^{1/2} \int \limits_{0}^{+\infty} d x F(x),\non \\
&F(x) = \frac{1}{x^{3/2}}  \left[ \frac{(\hat{a}^2+1)^{3/2}}{(x^2/3 - \hat{a} x + \hat{a}^2+1)^{3/2}} - 1  \right],
\end{align}
where we rescaled the variables as $x=\gamma s$ and $\hat{a}=\gamma a$. The angular distribution of the radiation power, according to general expression (\ref{W}) in $2+1$ dimensions will be 
\begin{equation}
\label{eq:2+1_W_gen}
\frac{dW_{2+1}}{d\phi} = R \, T_{0i}^{\rm rad} n_{i},
\end{equation}
where  the unit vector directed to the observation point is $\bold{n}= [ \cos \phi; \sin \phi ] $. From the Eqs. $\eqref{eq:EMT_2+1_gen}, \eqref{eq:2+1_rho}$ and $\eqref{eq:2+1_c}$ we obtain the angular distribution of the radiation power
\begin{equation}
\label{eq:2+1_W_start}
\frac{dW_{2+1}}{d\phi} = \frac{g^2 \omega_{0} \gamma \J^2}{4 \pi (1-v \cos a)^3}.
\end{equation}
Passing to an angular integration variable $a$ instead of $\phi$ via $\eqref{eq:phi-a_relation}$ and taking into account the leading-$\gamma$ behaviour:
\begin{equation}
\frac{d a}{d \phi} \approx - \frac{\gamma^2}{\hat{a}^2 + 1},
\end{equation}
we come to the leading $\gamma$-approximation for the total synchrotron scalar radiation power
\begin{equation}
\label{eq:2+1_W_rescaled}
W_{2+1} = \frac{g^2 \omega_{0} \gamma^2}{4 \pi} \int \limits_{-\infty}^{+\infty} d \hat{a} \frac{\J^2}{(\hat{a}^2 + 1)^2}.
\end{equation}
The integral here is nothing but a numerical factor independent of any physical parameters.

Now we can show that the divergences in the Eq. (\ref{eq:2+1_W_rescaled}) coming from the lower integration point in $\J$ mutually cancel. To demonstrate this, we have to integrate twice the first term in the integral (\ref{eq:2+1_rad_int_lead}) by parts.  As a result we arrive at the following convergent integral:
\begin{equation}
\label{eq:2+1_reg_nm_fact}
\int \limits_{0}^{+\infty} d x F(x) = \int \limits_{0}^{+\infty} dx \frac{x^{1/2}}{(x^2/3-\hat{a}x+\hat{a}^2+1)^{5/2}} \left \lbrace 4 - \frac{15 (2x/3-\hat{a})^{2}}{x^2/3-\hat{a}x+\hat{a}^2+1} \right \rbrace.
\end{equation}
Numerical integration in (\ref{eq:2+1_W_rescaled}) gives the value $4\pi/\sqrt{3}$ up to the 5 digits for the integral in $\eqref{eq:2+1_W_rescaled}$. As a result, we find the final expression for the  scalar $(2+1)$-dimensional synchrotron radiation power:
\begin{equation}
\label{eq:2+1_W_result}
W_{2+1} = \frac{g^2 \omega_{0} \gamma^2}{\sqrt{3}}.
\end{equation}

\subsection{Spectral decomposition}
To double check the validity of the above calculation, we now calculate the spectral-angular distribution of total radiated energy using the Fourier decomposition $\eqref{eq:gen_sp-ang_en_dist}$. Taking into account the beaming of radiation in the instantaneous direction of the particle's velocity 
and the fact that the contribution of the particle history, coming from the tail part of the Green function   is limited by the proper time interval $\delta s \sim 1/ \gamma$, we understand that  the instantaneous intensity of radiation in a given direction is determined by the short arc of  the  circular trajectory of the order of $\delta l \sim R_0 / \gamma$ (see, for example, \cite{Baier}).
In view of this we can simplify the spectral-angular distribution of radiated energy $\eqref{eq:gen_sp-ang_en_dist}$ starting with
\begin{equation}
\frac{dP_0}{d\omega d\Omega} = \frac{\omega g^2}{4 \pi \gamma^2} \int \limits_{-\infty}^{+\infty} d t_1 \int \limits_{-\infty}^{+\infty} d t_2 \, e^{i \omega (t_1 - t_2) - i \bold{k} (\bold{z}(t_1) - \bold{z}(t_2))},
\end{equation}
where we put $n=2$ and used the formula $\eqref{eq:sphere_area}$ for $n=2$. Transforming the integration variables as
\be \label{tt}
t_1 = t - t_0/2,\qquad
t_2 = t + t_0/2,
\ee
we obtain the instantaneous spectral-angular distribution of the radiation power:
\begin{equation}
\label{eq;2+1_W_sp-ang_distr}
\frac{dW}{d\omega d\Omega} = \frac{dP_0 / dt}{d\omega d\Omega} = \frac{\omega g^2}{4 \pi \gamma^2} \int \limits_{-\infty}^{+\infty} d t_0 \, e^{ - i \omega t_0 - i \bold{k} (\bold{z}(t - t_0 / 2) - \bold{z}(t + t_0 / 2))}.
\end{equation}

In the ultrarelativistic case, the main contribution to this integral is given by an interval $\delta t_0 \sim 1/\gamma$ near $t_0 = 0$.  Expanding   the exponent   in the Taylor series at $t_0 = 0$, we find to the leading order in $\gamma$
\begin{equation}
- i \omega t_0 - i \bold{k} (\bold{z}(t - t_0 / 2) - \bold{z}(t + t_0 / 2)) \sim - i \omega t_0 \left( \frac{1}{2} \left( a^2 + \frac{1}{\gamma^2} \right) + \frac{\omega_0^2 t_0^2}{24} \right),
\end{equation}
where $\bold{k}=\omega [ \cos \phi ; \sin \phi ] $,  $\bold{z} (t) = R_0 [ \cos \omega_0 t ; \sin \omega_0 t ] $ and we introduced an angular variable $a=\omega_0 t - \phi + \pi/2$. Then the leading-$\gamma$ asymptotic of the spectral distribution of radiation intensity takes the form:
\begin{equation}
\frac{dW_{2+1}}{d\omega} = \frac{\omega g^2}{4 \pi \gamma^2} \int \limits_{0}^{2 \pi} d\phi \int \limits_{-\infty}^{+\infty} dt_0 \exp \left \lbrace - i \omega t_0 \left( \frac{1}{2} \left( a^2 + 1/\gamma^2 \right) + \frac{\omega_0^2 t_0^2}{24} \right) \right \rbrace.
\end{equation}
Similarly to the four-dimensional theory  \cite{Baier}, after rescaling of the integration variables $\hat{a}=a\gamma$ and $t'=\frac{1}{2} \left( \omega/\omega_0 \right)^{1/3} \omega_0 t_0$, one can express the integral through the Airy function
 \cite{abr}
\begin{equation}
\label{eq:Airy_definition}
\Ai (u) = \frac{1}{2\pi} \int \limits_{-\infty}^{+\infty} dt' \exp \left \lbrace i \left( ut' +\frac{t'^3}{3} \right) \right \rbrace
\end{equation}
as follows
\begin{equation}
\label{eq:2+1_sp_intens_distr_Airy}
\frac{dW_{2+1}}{d\omega} = \left( \frac{\omega}{\omega_0} \right)^{2/3} \frac{g^2}{\gamma^3} \int \limits_{-\infty}^{+\infty} d\hat{a} \Ai \left( \left( \frac{\omega}{\omega_0 \gamma^3} \right)^{2/3} \left( \hat{a}^2 + 1 \right) \right).
\end{equation}
Now introduce $s=x\left( \hat{a}^2 + 1 \right)$, where $x=\left( \omega / \omega_0 \gamma^3 \right)^{2/3}$, 
and using the formula
  \cite{olivier2010airy}
\begin{equation}
\int \limits_{x}^{+\infty} ds \frac{\Ai (s)}{(s-x)^{1/2}} = 2^{2/3} \pi \Ai^2 \left( \frac{x}{2^{2/3}} \right),
\end{equation}
we perform integration in $\eqref{eq:2+1_sp_intens_distr_Airy}$ 
arriving at the spectral distribution
\begin{equation}
\label{eq:2+1_dW/dw}
\frac{dW_{2+1}}{d\omega} = \left( \frac{\omega}{\omega_0} \right)^{2/3} \frac{g^2}{x^{1/2} \gamma^3} 2^{2/3} \pi \Ai^2 \left( \frac{x}{2^{2/3}} \right).
\end{equation}

We can integrate $\eqref{eq:2+1_dW/dw}$ over the spectrum passing to an integration variable  $x=\left( \omega / \omega_0 \gamma^3 \right)^{2/3}$, leading to  
\begin{equation}
W_{2+1} = \frac{3}{2} 2^{2/3} \pi g^2 \omega_0 \gamma^2 \int \limits_{0}^{+\infty} dx \, x \Ai^2 \left( \frac{x}{2^{2/3}} \right).
\end{equation}
Finally, using  the integral of the squared Airy function \cite{olivier2010airy},
\begin{equation}
\int \limits_{0}^{+\infty} ds \, s \Ai^2 (s) = \frac{1}{6 \sqrt{3} \pi},
\end{equation}
we obtain the total radiation power 
\begin{equation}
\label{eq:2+1_W_result_spectral}
W_{2+1} = \frac{g^2 \omega_0 \gamma^2}{\sqrt{3}},
\end{equation}
coinciding with  $\eqref{eq:2+1_W_result}$.  
\section{$\bold{(4+1)}$ theory}
Calculations in the $(4+1)$-dimensional theory  conceptually are the same, so we just briefly describe the main steps.
\subsection{Coordinate representation}
The retarded solution of the wave equation   $\eqref{eq:sc_eq_motion}$ reads:
\begin{eqnarray}
\label{eq:4+1_gen_sol}
\varphi_{\rm ret}^{4+1} (x) = -2\pi^2 \int G_{\rm ret}^{4+1} (x-x') j(x') d^{4+1} x', \\
\label{eq:4+1_gr_fn}
G_{\rm ret}^{4+1} (X) = \frac{\theta (X^0)}{2\pi^2} \left \lbrace \frac{\delta (X^2)}{(X^2)^{1/2}} - \frac{1}{2} \frac{\theta (X^2)}{(X^2)^{3/2}} \right \rbrace.
\end{eqnarray}
The corresponding field gradient takes the following form:
\begin{equation}
\label{eq:4+1_gen_grad}
 \varphi_\mu^{\rm ret}(x) = - g \int d \tau \theta (X^0 (z)) \left \lbrace \frac{3}{2} \frac{\theta (X^2 (z))}{(X^2 (z))^{5/2}} + 2 \frac{\delta' (X^2 (z))}{(X^2 (z))^{1/2}} - 2 \frac{\delta (X^2 (z))}{(X^2 (z))^{3/2}} \right \rbrace,
\end{equation}
where $\delta' (x) = d \delta (x) / dx$.  
Using the relation
\begin{equation}
\frac{d X^2 (\tau)}{d \tau} = - 2 \left(v (\tau) X (z)\right),
\end{equation}
we can integrate the $\delta'$-term by parts,
arriving at
\begin{multline}
\varphi_\mu^{\rm ret}(x) = - g \int \limits_{-\infty}^{\hat{\tau}} d \tau \frac{1}{2(X^2 (z))^{1/2}} \left \lbrace \frac{3}{(X^2 (z))^{2}} X_{\mu} (z) - \frac{\delta (\tau - \hat{\tau})}{\hat{\rho}X^2 (z)} X_{\mu} (z) - \right. \\ \left. - \frac{\delta (\tau - \hat{\tau})}{\hat{\rho} (v (\tau) X (z))^2} (a (\tau) X (z) - 1) X_{\mu} (z) - \frac{\delta (\tau -\hat{\tau})}{\hat{\rho} v (\tau) X (z)} v_{\mu} (\tau) \right \rbrace,
\end{multline}
where we have used the relation $\eqref{eq:d-t_rel}$ and introduced an acceleration five-vector $a^{\mu} (\tau) = d^2 z^{\mu} (\tau)/d \tau^2$.  Expanding this in terms of half-integer powers of the inverse distance $1/\rh$ we obtain the radiated field
\begin{equation}
\label{eq:4+1_rad_grad}
\varphi_{\mu}^{\rm rad} (x) = - \frac{g \hat{c}_{\mu}}{2^{5/2}\hat{\rho}^{3/2}} \int \limits_{-\infty}^{\hat{\tau}} d \tau \left \lbrace \frac{3}{2} \frac{1}{(\Z\hat{c})^{5/2}} - \frac{\delta (\tau - \hat{\tau})}{(\Z\hat{c})^{3/2}} - \frac{2a\hat{c} \, \delta (\tau - \hat{\tau})}{(\Z\hat{c})^{1/2} (v\hat{c})^2} \right \rbrace.
\end{equation}
One can  simplify this expression taking into account the relations   $\eqref{eq:ret_q_1}$ and  transformations  similar to $\eqref{eq:delta_transform}$ and $\eqref{eq:2+1_st_contr}$:
\begin{eqnarray}
\int \limits_{-\infty}^{\hat{\tau}} d \tau \frac{\delta (\tau - \hat{\tau})}{(\Z\hat{c})^{3/2}} = \frac{3}{2} \int \limits_{-\infty}^{\hat{\tau}} \frac{d\tau}{(\hat{\tau}-\tau)^{5/2}}, \\
\int \limits_{-\infty}^{\hat{\tau}} d \tau \frac{2a\hat{c} \, \delta (\tau - \hat{\tau})}{(\Z\hat{c})^{1/2} (v\hat{c})^2} = \int \limits_{-\infty}^{\hat{\tau}} d \tau \frac{\hat{a}\hat{c}}{(\hat{\tau}-\tau)^{3/2}}.
\end{eqnarray}
This gives:
\begin{equation}
\label{eq:4+1_rad_sc_gr}
\varphi_{\mu}^{\rm rad} (x) = - \frac{g \hat{c}_{\mu}}{2^{5/2}\hat{\rho}^{3/2}} \int \limits_{-\infty}^{\hat{\tau}} d \tau \left \lbrace \frac{3}{2} \frac{1}{(\Z\hat{c})^{5/2}} - \frac{3}{2} \frac{1}{(\hat{\tau}-\tau)^{5/2}} - \frac{\hat{a}\hat{c}}{(\hat{\tau}-\tau)^{3/2}} \right \rbrace.
\end{equation}
Now we have the sum of three integrals whose divergences at the upper limit mutually cancel: the second term is the counterterm for the first eliminating its leading divergence, while the third one is the counterterm eliminating the remaining divergence.
 
The $(4+1)$-dimensional  on shell energy-momentum tensor evaluated with the radiated part of the field, with account for relations  $\eqref{eq:ret_q_1}$, will read:
\begin{equation}
T_{\mu\nu}^{\rm rad} (x) = \frac{g^2 \hat{c}_{\mu} \hat{c}_{\nu}}{64 \pi^2 \hat{\rho}^3} \J^{2} (x),
\end{equation}
where  the $(4+1)$-dimensional integral radiation amplitude is
\begin{equation}
\label{eq:4+1_rad_int_start}
\J (x) = \int \limits_{-\infty}^{\hat{\tau}} d \tau \left \lbrace \frac{3}{2} \frac{1}{(\Z\hat{c})^{5/2}} - \frac{3}{2} \frac{1}{(\hat{\tau}-\tau)^{5/2}} - \frac{\hat{a}\hat{c}}{(\hat{\tau}-\tau)^{3/2}} \right \rbrace.
\end{equation}
Again, the energy-momentum tensor obtained satisfies all Teitelboim's requirements and therefore describes the radiated energy-momentum indeed.
\subsection{Synchrotron radiation}
Assuming the wordline to lie in the equatorial plane,  
\begin{equation}
\label{eq:4+1_worldline}
z^{\mu} (\tau) = [ \gamma \tau , R_0 \cos \omega_0 \gamma \tau , R_0 \sin \omega_0 \gamma \tau , 0 , 0 ],
\end{equation}
and using for the retarded proper time  
$\hat{\tau} = {(t-R)}/{\gamma}$,
 we find
\begin{align}
\label{eq:4+1_rho}
&\hat{\rho} = \gamma R \left( 1 + v \sin (\omega_0 \gamma \hat{\tau} - \phi) \cos \theta \cos \zeta \right), \\
\label{eq:4+1_c}
&\hat{c}^{\mu} = \frac{R}{\hat{\rho}} [ 1 , \cos \phi \cos \theta \cos \zeta , \sin \phi \cos \theta \cos \zeta , - \sin \theta \cos \zeta , - \sin \zeta ],
\end{align}
where we used the hyperspherical coordinates for the observation point
\begin{equation}
x^{\mu} = [ t , R \cos \phi \cos \theta \cos \zeta , R \sin \phi \cos \theta \cos \zeta , - R \sin \theta \cos \zeta , - R \sin \zeta ].
\end{equation}
Calculating the contractions $\Z\hat{c}$ and $\hat{a}\hat{c}$, we then pass to the integration variable  $s=\omega_0 \gamma (\hat{\tau}-\tau)$ and introduce the angular  variable $a = \omega_0 \gamma \hat{\tau} - \phi +\frac{\pi}{2}$. Zero values of these variables correspond to the particle's instant direction of motion at the retarded moment of proper time $\hat{\tau}$. After  some algebra, we arrive at the following representation of the integral amplitude of radiation:
\begin{align}
\label{eq:4+1_rad_int_trans}
&\J (x) = (\omega_0\gamma)^{3/2} \int \limits_{0}^{+\infty} ds \, G(s), \\
&G(s) = \frac{3}{2} \left( \left \lbrack \frac{\delta \Delta}{ s - v \Delta \left( \sin a + \sin(s-a) \right)} \right \rbrack^{5/2} - \frac{1}{s^{5/2}} \right) - \frac{v \Delta \sin a}{s^{3/2} \left( 1 - v \Delta \cos a \right)},
\end{align}
where we denoted $\Delta = \cos \theta \cos \zeta$ and $\delta = 1 - v \cos a$.
One can see  that in the ultrarelativistic case  $v \to 1$ the main part of the energy is radiated in a narrow cone around the instantaneous direction of  velocity   $a=0$, $\theta=0$, $\zeta=0$ within $\delta a \sim 1/\gamma$, $\delta \theta \sim 1/\gamma$, $\delta \zeta \sim 1/\gamma$.
Using the leading-$\gamma$ relation between the angular variables following from the equation for $\hat{\tau}$ $\eqref{eq:ret_prop_time_eq}$,
\begin{equation}
\label{eq:4+1_phi-a_rel}
\frac{da}{d\phi} = - \frac{1}{1 - v \Delta \cos a} \approx - \frac{1}{2 \gamma^2} \left( \hat{a}^2 + \hat{\theta}^2 + \hat{\zeta}^2 + 1 \right),
\end{equation}
and rescaling the integration variables $\lbrace s, a, \theta, \zeta \rbrace \to \lbrace x=s\gamma, \hat{a}=a\gamma, \hat{\theta}=\theta\gamma, \hat{\zeta}=\zeta\gamma \rbrace$, we obtain from the Eq. $\eqref{eq:4+1_rad_int_trans}$ in the leading order:
\begin{equation}
\label{eq:4+1_rad_int_lead}
G(x) = \frac{\gamma^{5/2}}{x^{3/2}} \left \lbrace \frac{3}{2x} \left( \left \lbrack \frac{ \hat{A} }{ x^2/3 - \hat{a}x + \hat{A} } \right \rbrack^{5/2} - 1 \right) - \frac{2\hat{a}}{ \hat{A} } \right \rbrace,
\end{equation}
where we denoted $\hat{A}=\hat{a}^2 + \hat{\theta}^2 +\hat{\zeta}^2 + 1$.

Divergences in the different terms of the radiation integral $\eqref{eq:4+1_rad_int_lead}$ at $s \sim 0$ at the upper limit are cancelled after the triple integration by parts of the first term in the integrand. Doing these integrations one can use  different regularisation parameters $\varepsilon = +0$    absorbing in them some numerical factors.

The total radiation power now will be given by the integral:
\begin{equation}
\label{eq:4+1_W_gen}
W_{4+1} = \int d\phi d\theta d\zeta \, R^3 \sin \theta \sin^2 \zeta \, T_{0i}^{\rm rad} n_{i}.
\end{equation}
After triple integration by parts in the radiation amplitude, the angular distribution of the radiation power will read
\begin{equation}
\label{eq:4+1_dW_result_nonfactor}
\frac{dW_{4+1}}{d\hat{a} d\hat{\theta} d\hat{\zeta}} = \frac{g^2 \omega_{0}^3 \gamma^6}{4 \pi^2} \hat{A} \left \lbrace \int \limits_{0}^{+\infty} dx \frac{ 315 x^{1/2} \Phi}{2 \Lambda^{9/2}} \left( \frac{\Phi^2}{\Lambda} - \frac{140}{315} \right) \right \rbrace^2,
\end{equation}
where we denoted $\Phi = 2x/3 - \hat{a}$ and $\Lambda = x^2/3 - \hat{a}x + \hat{A}$. 
In the leading-$\gamma$ approximation,the angular distribution is beamed, this is used similarly to the previous section. 
Integration over the angles is relegated to Appendix. The resulting factor is obtained by numerical integration after some analytic transformations and has  the value  $1/\sqrt{27}$ up to $5$ digits. The total radiation power of the $(4+1)$-dimensional scalar synchrotron radiation  will read:
\begin{equation}
\label{eq:4+1_W_result}
W_{4+1} = \frac{g^2 \omega_{0}^3 \gamma^6}{\sqrt{27}}.
\end{equation}
\subsection{Spectral decomposition}
Now from $\eqref{eq:gen_sp-ang_en_dist}$ we get:\be
\frac{dP_0}{d\omega d\Omega} = \frac{\omega^3 g^2}{16 \pi^2 \gamma^2} \int \limits_{-\infty}^{+\infty} dt_1 \int \limits_{-\infty}^{+\infty} dt_2 \, e^{i \omega (t_1 - t_2) - i \bold{k}(\bold{z}(t_1) - \bold{z}(t_2))}.
\end{equation}
After the transformation (\ref{tt}), we present 
the radiation power as
\begin{equation}
\label{eq:4+1_synch_gen}
\frac{dW }{d\omega d\Omega} = \frac{dP_0 / dt}{d\omega d\Omega} = \frac{\omega^3 g^2}{16 \pi^2 \gamma^2} \int \limits_{-\infty}^{+\infty} dt_0 e^{ - i \omega t_0 - \bold{k} ( \bold{z}(t - t_0/2) - \bold{z}(t + t_0/2) ) }.
\end{equation}
The main contribution to the integral comes from the region $\delta t_0 \sim 1/\gamma$ near $t_0 = 0$.  Thus  one can  
expand an expression in the  exponent of the integrand  into Taylor series at $t_0 = 0$ up to the leading order in $\gamma$ as
\begin{equation}
- i \omega t_0 - i \bold{k} (\bold{z}(t - t_0 / 2) - \bold{z}(t + t_0 / 2)) = - \frac{i \omega t_0}{2\gamma^2} \left( \hat{a}^2 + \hat{\theta}^2 + \hat{\zeta}^2 + 1 + \frac{\omega_0^2 t_0^2 \gamma^2}{12} \right)
\end{equation}
where  the angular variables are chosen in accordance with  the wave zone  calculations. We also rescaled the  angular variables  multiplying on $\gamma$ to stretch the limits $\hat{a}$, $\hat{\theta}$, $\hat{\zeta} \in (-\infty,+\infty)$. After that we arrive at the spectral distribution of the radiation power in the form:
\begin{equation}
\frac{dW_{4+1}}{d\omega} = \frac{\omega^3 g^2}{16 \pi^2 \gamma^5} \int \limits_{{\mathbb R}^3}  d\hat{a}   d\hat{\theta}   d\hat{\zeta} \int \limits_{-\infty}^{+\infty} dt_0 \exp \left \lbrace -\frac{i \omega t_0}{2\gamma^2} \left( 1 + \hat{a}^2 + \hat{\theta}^2 + \hat{\zeta}^2 + \frac{\omega_0^2 t_0^2 \gamma^2}{12} \right) \right \rbrace.
\end{equation}
Using the spherical coordinates for integration over the rescaled angular variables
\be 
\hat{a} = \rho \cos \alpha \sin \beta,\quad \ 
\hat{\theta}  = \rho \sin \alpha \sin \beta,\quad
\hat{\zeta} = \rho \cos \beta, 
\ee
where $\rho \in \lbrack0;+\infty)$, $\alpha \in \lbrack0;2\pi)$ and $\beta \in \lbrack0;\pi\rbrack$, we find the following integral representation of the spectral distribution of radiation intensity
\begin{equation}
\frac{dW_{4+1}}{d\omega} = \frac{\omega^3 g^2}{4 \pi \gamma^5} \int \limits_{0}^{+\infty} d\rho \, \rho^2 \int \limits_{-\infty}^{+\infty} dt_0 \, \exp \left \lbrace - \frac{i \omega t_0}{2\gamma^2} \left( 1 + \rho^2 + \frac{\omega_0^2 t_0^2 \gamma^2}{12} \right) \right \rbrace.
\end{equation}
Integration over $t_0$ could be performed in terms of Airy function $\eqref{eq:Airy_definition}$:  
\begin{equation}
\frac{dW_{4+1}}{d\omega} = \frac{\omega^3 g^2}{\omega_0 \gamma^5} \left( \frac{\omega_0}{\omega} \right)^{1/3} \int \limits_{0}^{+\infty} d\rho \, \rho^2 \Ai \left( \left( \frac{\omega}{\gamma^3 \omega_0} \right)^{2/3} \left( \rho^2 + 1 \right) \right).
\end{equation}
We can rescale  an integration variable $r=\rho \left( \omega / \omega_0 \gamma^3 \right)^{1/3}$ and introduce  a dimensionless parameter $x=\left( \omega   \omega_0 \gamma^3 \right)^{2/3}$ to simplify this to  
\begin{equation}
\frac{dW_{4+1}}{d\omega} = \left( \frac{\omega_0}{\omega} \right)^{1/3} \frac{\omega^2 g^2}{\gamma^2} \int \limits_{0}^{+\infty} dr \, r^2 \Ai \left( r^2 + x \right).
\end{equation}
Integration over $r$ can be performed using the integral \cite{olivier2010airy}:
\begin{equation}
\label{eq:Airy_squared_int}
\int \limits_{0}^{+\infty} dr \, r^2 \Ai \left( r^2 + x \right) = \frac{\pi}{2^{2/3}} \left( \Ai' \left( x / 2^{2/3} \right) \right)^2 - \frac{\pi x}{2^{4/3}} \Ai^2 \left( x / 2^{2/3} \right).
\end{equation}
Passing to  new dimensionless parameter $s=\left( \omega / 2 \omega_0 \gamma^3 \right)^{2/3}$ we finally obtain:
\begin{equation}
\label{eq:4+1_spectral_distr}
\frac{dW_{4+1}}{d\omega} = 2\pi g^2 \gamma^3 \omega_0^2 s^{5/2} \left( \left( \Ai' \left( s \right) \right)^2 - s \Ai^2 (s) \right).
\end{equation}

The total   power of synchrotron radiation in $(4+1)$ dimensions will read
\begin{equation}
\label{eq:4+1_W_synch_int}
W_{4+1} = 6 \pi g^2 \omega_0^3 \gamma^6 \int \limits_{0}^{+\infty} ds \, s^3 \left( \left( \Ai' (s) \right)^2 - s \Ai^2 (s) \right).
\end{equation}
Using the integrals \cite{olivier2010airy}  
\begin{align}
\label{eq:Airy_int_1}
\int \limits_{0}^{+\infty} ds \, s^3 \left( \Ai' (s) \right)^2 &= \frac{5}{18\sqrt{3}\pi}, \nonumber\\
\label{eq:Airy_int_2}
\int \limits_{0}^{+\infty} ds \, s^4 \Ai^2 (s) &= \frac{2}{9\sqrt{3}\pi}.\nonumber
\end{align}
we find for the total power  
\begin{equation}
\label{eq:4+1_W_result_spectral}
W_{4+1} = \frac{g^2 \omega_0^3 \gamma^6}{\sqrt{27}},
\end{equation}
which coincides with the result of calculation in the wave zone $\eqref{eq:4+1_W_result}$.  
\subsection{Arbitrary D}
Combining $\eqref{eq:2+1_W_result}$ with $\eqref{eq:4+1_W_result_spectral}$ and with the known result for $D=4$ \cite{Barut:1974ch}, we can conjecture that the power of the scalar synchrotron radiation in $D=n+1$ dimensions is given by the following formula
\begin{equation}
\label{eq:n+1_W_result}
W^{\rm sc}_{n+1} = g^2 \left( \frac{\omega_0 \gamma^2}{\sqrt{3}} \right)^{n-1}.
\end{equation}
\section{Conclusion}
In this paper, we have shown that in spite of violation of the Huygens principle in odd dimensions, radiation from an accelerated charge can be computed integrating the flux of the energy-momentum in the asymptotic wave zone in a standard way. However, an important difference with the even-dimensional case is that the long range component of the retarded field gradient depends on the entire history of motion preceding the retarded proper time. Another unusual feature is that the amplitude is primarily presented by a sum of divergent terms, whose divergences mutually cancel, however. Integrating by parts, one is able to make the cancellation explicit and to obtain a finite result. 

We computed explicitly the synchrotron radiation power in $D=3,\,5$ dimensions showing that the integral over history (tail) is effectively localized in the relativistic limit, so the resulting expression can be found analytically. Moreover, we have found that, together with the previously known result for $D=4$ \cite{Barut:1974ch}, the power is described by the universal formula (\ref{eq:n+1_W_result}).
We conjecture that this might be valid in any $D$.

Our results were double checked by an independent spectral decomposition method based on expansion of all quantities in the Fourier integrals. This second way is closer to the quantum theory calculations, and it does not make difference between even and odd dimensions. Note that spectral-angular distributions have similar features in the ultrarelativistic case, in particular, beaming in odd and even dimensions
 
We now briefly discuss relation to other work. Our formula agrees with  Shuryak et al. \cite{Shuryak:2011tt} calculation of synchrotron radiation power in $2+1$ dimensions via radiation reaction, but disagrees with their result in $D=5$. We also disagree with the results by Yaremko also obtained via radiation reaction. We agree with qualitative estimates of the power by Mironov et. al. \cite{Mironov:2006wi} and with the results obtained by Cardoso et. al. \cite{Cardoso:2007uy} for even-dimensional spacetimes.

\section*{Acknowledgments}

This work of M.Kh. was supported by the "BASIS" Foundation grant 18-2-6-41-1. The
authors would like to acknowledge the networking support
of the COST Action No. CA16104.

\section*{Appendix}

The computation below belongs to Igor Bogush \cite{Bogush}. Consider the integral 
\begin{align}
J = \frac{1}{4\pi^2} \int \limits_{\mathbb{R}^3}  d\hat{a} d\hat{\theta} d\hat{\zeta} \, (\hat{A}+1) I^2, \\
I = \int \limits_{0}^{+\infty} dx \sqrt{x} \frac{f'}{f^{9/2}} \left( \frac{315}{2} \frac{f'^2}{f} - 70 \right), \\
f(x) = x^2/3 - ax + \hat{A} + 1,
\end{align}
where $\hat{A} = \hat{a}^2 + \hat{\theta}^2 + \hat{\zeta}^2$. Introducing cylindrical coordinates in the space of $(\hat{a},\hat{\theta},\hat{\zeta})$ by use of $b^2 = \hat{\theta}^2 + \hat{\zeta}^2$ we obtain:
\begin{align}
J = \frac{1}{2\pi} \int \limits_{-\infty}^{+\infty} d\hat{a} \int \limits_{0}^{+\infty} db \, b \left( \hat{a}^2 + b^2 +1 \right) I^2, \\
f(x) = \frac{3}{4} (2x/3 - \hat{a})^2 + \frac{1}{4} \hat{a}^2 + b^2 + 1.
\end{align}

Passing to the integration variable $y = 2x/3 - \hat{a}$ and denoting $c^2 = \hat{a}^2 + 4(b^2+1)$ we find
\begin{align}
f(x) = \frac{1}{4}(3y^2 + c^2), \quad f'(x)=y, \\
I = 70 \cdot 2^9 \left( \frac{3}{2} \right)^{3/2} \int \limits_{-\hat{a}}^{+\infty} dy \frac{y\sqrt{y+\hat{a}}}{(3y^2+c^2)^{9/2}} \left( \frac{9y^2}{3y^2+c^2} - 1 \right).
\end{align}
This integral can be evaluated in terms of the complete elliptic integrals $E(d)$ and $K(d)$
\begin{align}
I = \frac{1}{4 \cdot 3^{1/4} C} \left( A \left( 2 E(d) + K(d) \right) + B K(d) \right), \\
A = - 288\hat{a} (b^2+1)(\hat{a}^2+3b^2+3), \\
B = \sqrt{3} \sqrt{\hat{a}^2+b^2+1} \left( 28\hat{a}^4 + 44\hat{a}^2 (b^2+1) - 80 (b^2+1)^2 \right), \\
C = \left( \hat{a}^2 + 4(b^2+1) \right)^2 (\hat{a}^2 + b^2 + 1)^{15/4}, \\
d = \frac{1}{2} \left( \frac{\sqrt{3}\hat{a}}{2\sqrt{\hat{a}^2+b^2+1}} + 1 \right).
\end{align}
As a result, we obtain the amplitude of radiation intensity in the form of double integral
\begin{equation}
J = \frac{1}{32\sqrt{3}\pi} \int \limits_{-\infty}^{+\infty} d\hat{a} \int \limits_{0}^{+\infty} db \frac{b \left( A(2E(d)+K(d))+BK(d) \right)^2}{(\hat{a}^2+4(b^2+1))^{4}(\hat{a}^2+b^2+1)^{13/2}},
\end{equation}
which can be easily evaluated numerically, giving   $J=1/\sqrt{27}$ up to $5$ digits.


\begin{thebibliography}{4}

\bibitem{Rubakov}
    V.~A.~Rubakov,
    ``Large and infinite extra dimensions: An introduction,''
    Phys.\ Usp.\  {\bf 44}, 871 (2001)
    [Usp.\ Fiz.\ Nauk {\bf 171}, 913 (2001)]
    [arXiv:hep-ph/0104152].

\bibitem{Arefeva2014} 
    I.~Ya~Aref’eva,
    "Holographic approach to quark—gluon plasma in heavy ion collisions",
    Phys. Usp. 57 527–555 (2014).
    
\bibitem{Cardoso:2013vpa} 
  V.~Cardoso, R.~Emparan, D.~Mateos, P.~Pani and J.~V.~Rocha,
  ``Holographic collisions in confining theories,''
  JHEP {\bf 1401}, 138 (2014)
  [arXiv:1310.7590 [hep-th]].
  
\bibitem{Ehr}
    P.~Ehrenfest,
    ``How do the fundamental laws of physics make manifest that space has 3 dimensions?” 
    Ann. Phys. (Leipzig) 61, 440 (1920).

\bibitem{courant2008methods} 
    R.~Courant and D.~Hilbert,
    ``Methods of Mathematical Physics: Partial Differential Equations,''
    Wiley Classics Library
    (Wiley, 2008).
    
\bibitem{hadamard2014lectures}
	J.~Hadamard,
	``Lectures on Cauchy's Problem in Linear Partial Differential Equations,''
	(Dover Publications, 2014).
	
\bibitem{IvSo48} D.~Ivanenko and A.~Sokolov,
    Sov. Phys. Doklady, \textbf{36}, 37, (1940); ``Classical field
    theory'' (in Russian), Moscow, 1948; ``Klassische Feldtheorie'',
    Berlin, 1953.
    
\bibitem{Barvinsky:2003jf}
    A.~O.~Barvinsky and S.~N.~Solodukhin,
    ``Echoing the extra dimension,''
    Nucl. \ Phys. \ B {\bf 675} (2003) 159
    [arXiv:0307011 [hep-th]].
    
\bibitem{Deffayet:2007kf} 
    C.~Deffayet and K.~Menou,
    ``Probing Gravity with Spacetime Sirens,''
    Astrophys.\ J.\  {\bf 668}, L143 (2007)
    [arXiv:0709.0003 [astro-ph]].
    
\bibitem{Andriot:2017oaz}
    D.~Andriot and G.~L.~Gómez,
    ``Signatures of extra dimensions in gravitational waves,''
    JCAP {\bf 1706}, 048 (2017)
    Erratum: [JCAP {\bf 1905}, E01 (2019)]
    [arXiv:1704.07392 [hep-th]].
    
\bibitem{Chakravarti:2019aup} 
    K.~Chakravarti, S.~Chakraborty, K.~S.~Phukon, S.~Bose and S.~SenGupta,
    ``Constraining extra-spatial dimensions with observations of GW170817,''
    [arXiv:1903.10159 [gr-qc]].
    
\bibitem{Yu:2019jlb}
    H.~Yu, Z.~C.~Lin and Y.~X.~Liu,
    ``Gravitational waves and extra dimensions: a short review,''
    Commun.\ Theor.\ Phys.\  {\bf 71} (2019) no.8,  991
    [arXiv:1905.10614 [gr-qc]].
    
\bibitem{Kwon:2019gsa}
    O.~K.~Kwon, S.~Lee and D.~D.~Tolla,
    ``Gravitational Waves as a Probe of the Extra Dimension,''
    Phys.\ Rev.\ D {\bf 100} (2019) 084050
    [arXiv:1906.11652 [hep-th]].
    
\bibitem{Cardoso:2019vof} 
    V.~Cardoso, L.~Gualtieri and C.~J.~Moore,
    ``Gravitational waves and higher dimensions: Love numbers and Kaluza-Klein excitations,''
    Phys.\ Rev.\ D {\bf 100}, no. 12, 124037 (2019)
    [arXiv:1910.09557 [gr-qc]].
    
\bibitem{Vagnozzi:2019apd} 
    S.~Vagnozzi and L.~Visinelli,
    ``Hunting for extra dimensions in the shadow of M87*,''
    Phys.\ Rev.\ D {\bf 100}, no. 2, 024020 (2019)
    [arXiv:1905.12421 [gr-qc]].
    
\bibitem{Galtsov:2012pcw} 
    D.~Gal'tsov, P.~Spirin and T.~N.~Tomaras,
    ``Gravitational bremsstrahlung in ultra-planckian collisions,''
    JHEP {\bf 1301}, 087 (2013)
    [arXiv:1210.6976 [hep-th]].
    
\bibitem{Berti:2010gx} 
    E.~Berti, V.~Cardoso and B.~Kipapa,
    ``Up to eleven: radiation from particles with arbitrary energy falling into higher-dimensional black holes,''
    Phys.\ Rev.\ D {\bf 83}, 084018 (2011)
    [arXiv:1010.3874 [gr-qc]].
    
\bibitem{Galtsov:2010vtu} 
    D.~V.~Galtsov, G.~Kofinas, P.~Spirin and T.~N.~Tomaras,
    ``Classical ultrarelativistic bremsstrahlung in extra dimensions,''
    JHEP {\bf 1005}, 055 (2010)
    [arXiv:1003.2982 [hep-th]].
    
\bibitem{Frolov:2003mc}
    V.~Frolov, M.~Snajdr and D.~Stojkovic,
    ``Interaction of a brane with a moving bulk black hole,''
    Phys. \ Rev. \ D {\bf 68} (2003) 044022
    [arXiv:gr-qc/0304083].
    
\bibitem{Galtsov:2015yyr} 
    D.~Gal'tsov, E.~Melkumova and P.~Spirin,
    ``Branestrahlung: radiation in the particle-brane collision,''
    Phys.\ Rev.\ D {\bf 93}, no. 4, 045018 (2016)
    [arXiv:1512.04607 [hep-th]].
    
\bibitem{Galtsov:2017udh} 
    D.~Gal'tsov, E.~Melkumova and P.~Spirin,
    ``Piercing of domain walls: new mechanism of gravitational radiation,''
    JHEP {\bf 1801}, 120 (2018)
    [arXiv:1711.01114 [hep-th]].
    
\bibitem{Kosyakov1999}
    B.~P.~Kosyakov,
    ``Exact solutions of classical electrodynamics and the Yang-Mills-Wong theory in even-dimensional space-time,''
    Theor.\ Math.\ Phys.\ {\bf 119} (1999) no. 1, 493.
    
\bibitem{Kosyakov:1992}
	B.~P.~Kosyakov,
	``Radiation in electrodynamics and in Yang-Mills theory,''
	Phys.\ Usp.\ {\bf 35} (1992) no. 2, 135.
	
\bibitem{Cardoso:2002pa}
	V.~Cardoso, O.~J.~C.~Dias and J.~P.~S.~Lemos,
	``Gravitational radiation in D-dimensional space-times,''
	Phys. \ Rev. \ D {\bf 67} (2003) 064026
	[arXiv:0212168 [hep-th]].
	
\bibitem{Cardoso:2007uy}
	V.~Cardoso, M.~Cavaglia and J.~Q.~Guo,
	``Gravitational Larmor formula in higher dimensions,''
	Phys. \ Rev. \ D {\bf 75} (2007) 084020
	[arXiv:0702138 [hep-th]].
	
\bibitem{Gurses:2003cc} 
    M.~Gurses and O.~Sarioglu,
    ``Lienard-Wiechert potentials in even dimensions,''
    J.\ Math.\ Phys.\  {\bf 44}, 4672 (2003)
    [arXiv:hep-th/0303078].
    
\bibitem{Yar7}
    Y.~Yaremko,
    ``Radiation reaction in 2+1 electrodynamics,''
    J. \ Math. \ Phys. \ {\bf 48} (2007) no. 9, 092901
    [arXiv:0907.3060 [physics.class-ph]].
    
\bibitem{Yar12}
    Y.~Yaremko,
    ``Renormalization and radiation reaction in 2+1 electrodynamics,''
    [arXiv:1207.5153 [math-ph]].
    
\bibitem{Yaremko:2007zz}
    Y.~Yaremko,
    ``Self-force in 2+1 electrodynamics,''
    J. \ Phys. \ A {\bf 40} (2007) 13161
    [arXiv:0907.3029 [physics.class-ph]].
    
\bibitem{Shuryak:2011tt}
    E.~Shuryak, {H.-U.}~Yee and I.~Zahed,
    ``Self-force and synchrotron radiation in odd space-time dimensions,''
    Phys. \ Rev. \ D {\bf 85} (2012) 104007
    [arXiv:1111.3894 [hep-th]].
    
\bibitem{Kazinski:2002mp}
    P.~O.~Kazinski, S.~L.~Lyakhovich and A.~A.~Sharapov,
    ``Radiation reaction and renormalization in classical electrodynamics of point particle in any dimension,''
    Phys. \ Rev. \ D {\bf 66} (2002) 025017
    [arXiv:0201046 [hep-th]].
    
\bibitem{Galtsov:2001iv}
    D.~V.~Galtsov,
    ``Radiation reaction in various dimensions,''
    Phys. \ Rev. \ D {\bf 66} (2002) 025016
    [arXiv:0112110 [hep-th]].
    
\bibitem{Dai:2013cwa}
    {D.-C.}~Dai and D.~Stojkovic,
    ``Origin of the tail in Green's functions in odd-dimensional space-times,''
    Eur. \ Phys. \ J. \ Plus \ {\bf 128} (2013) 122
    [arXiv: [hep-th]].
    
\bibitem{Kosyakov:2018wek}
    B.~P.~Kosyakov,
    ``Self-interaction in classical gauge theories and gravitation,''
    Phys.\ Rept.\ {\bf 812} (2019) 1
    [arXiv:1812.03290 [hep-th]].
    
\bibitem{Kosyakov:51252}
    B.~P.~Kosyakov,
    ``Introduction to the Classical Theory of Particles and Fields''
    (Springer, Berlin, Heidelberg, 2007).
    
\bibitem{Dirac}
    P.~A.~M.~Dirac,
    ``Classical theory of radiating electrons,''
    Proc.\ R.\ Soc.\ Lond.\ A {\bf 167} (1938) no. 929, 148.
    
\bibitem{Kazinski:2005gx}
	P.~O.~Kazinski and A.~A.~Sharapov,
	``Radiation back-reaction and renormalization in classical field theory with singular sources,''
	Theor. \ Math. \ Phys. \ {\bf 143} (2005) 798.
	
\bibitem{Shilov} 
    G.~E.~Shilov,
    ``Generalized functions and partial differential equations,''
    (Gordon and Breach, New York, 1968).
    
\bibitem{DeWitt:1960fc}
	B.~S.~DeWitt and R.~W.~Brehme,
	``Radiation damping in a gravitational field,''
	Annals \ Phys. \ {\bf 9} (1960) 220.
	
\bibitem{Barack:2018yvs}
	L.~Barack and A.~Pound,
	``Self-force and radiation reaction in general relativity,''
	Rept. \ Prog. \ Phys. \ {\bf 82} (2019) 016904
	[arXiv:1805.10385 [gr-qc]].
	
\bibitem{Galtsov:2007zz}
    D.~V.~Gal'tsov and P.~A.~Spirin,
    ``Radiation reaction in curved even-dimensional spacetime,''
    Grav. \ Cosmol. \ {\bf 13} (2007) 241
    [arXiv:1012.3085 [gr-qc]].
	
\bibitem{Spirin}
    P.~A.~Spirin,
    ``Massless field emission in the space-time of extra dimensions,''
    Grav.\ Cosmol.\ {\bf 15} (2009) no. 1, 82.
    
\bibitem{LL}
    L.~D.~Landau and E.~M.~Lifshitz,
    ``The Classical Theory of Fields''
    (Elsevier, 2013).
    
\bibitem{Rohrlich1961}
    F.~Rohrlich,
    ``The definition of electromagnetic radiation,''
    Il Nuovo Cimento (1955-1965)\ {\bf 21} (1961) no. 5, 811.
    
\bibitem{rohr}
    F.~Rohrlich,
    ``Classical Charged Particles''
    (World Scientific, 2007).
    
\bibitem{Teit}
	C.~Teitelboim,
	``Splitting of the Maxwell Tensor: Radiation Reaction without Advanced Fields,''
	Phys.\ Rev.\ D {\bf 1} (1970) 1572.
	
\bibitem{Galtsov:2004uqu}
    D.~V.~Gal'tsov and P.~Spirin,
    ``Radiation reaction reexamined: Bound momentum and Schott term,''
    Grav.\ Cosmol.\ {\bf 12} (2006) 1
    [arXiv:0405121 [hep-th]].
    
\bibitem{Galtsov:2010tny}
    D.~V.~Gal'tsov,
    ``Radiation Reaction and Energy-Momentum Conservation,''
    Fundam.\ Theor.\ Phys.\ {\bf 162} (2011) 367
    [arXiv:1012.2846 [gr-qc]].
    
\bibitem{Birnholtz:2015hua} 
    O.~Birnholtz and S.~Hadar,
    ``Gravitational radiation-reaction in arbitrary dimension,''
    Phys.\ Rev.\ D {\bf 91} (2015), no. 12, 124065
    [arXiv:1501.06524 [gr-qc]].
    
\bibitem{Harte:2018iim} 
    A.~I.~Harte, P.~Taylor and É.~É.~Flanagan,
    ``Foundations of the self-force problem in arbitrary dimensions,''
    Phys.\ Rev.\ D {\bf 97} (2018), no. 12, 124053
    [arXiv:1804.03702 [gr-qc]].
    
\bibitem{Birnholtz:2013ffa}
    O.~Birnholtz and S.~Hadar,
    ``Action for reaction in general dimension,''
    Phys.\ Rev.\ D {\bf 89} (2014) no. 4, 045003
    [arXiv:1311.3196 [hep-th]].
    
\bibitem{Porto:2016pyg} 
    R.~A.~Porto,
    ``The effective field theorist’s approach to gravitational dynamics,''
    Phys.\ Rept.\  {\bf 633}, 1 (2016)
    [arXiv:1601.04914 [hep-th]].
    
 
\bibitem{Breuer:1974uc}
	R.~A.~Breuer, P.~L.~Chrzanowksi, H.~G.~Hughes and C.~W.~Misner,
	``Geodesic synchrotron radiation,''
	Phys.\ Rev.\ D {\bf 8} (1973) 4309.
	
\bibitem{2009arXiv0901.3488L}
    R.~R.~Landim,
    ``On the Laplace equation in d-dimension,''
    [arXiv:0901.3488 [math-ph]].
	
\bibitem{zwillinger2014table}
	D.~Zwillinger,
	``Table of Integrals, Series, and Products''
	(Elsevier Science, 2014).
	
\bibitem{watson}
	G.~N.~Watson,
	``A Treatise on the Theory of Bessel Functions''
	Cambridge Mathematical Library
	(Cambridge University Press, 1995).
    
\bibitem{Baier}
    V.~N.~Baier, V.~M.~Katkov and V.~S.~Fadin,
    ``Radiation of relativistic electrons'' (in Russian)
    (1973).
    
\bibitem{abr}
    M.~Abramowitz and I.~A.~Stegun,
    ``Handbook of Mathematical Functions with Formulas, Graphs, and Mathematical Tables''
    (Dover, New York, 1964).
    
\bibitem{olivier2010airy}
	V.~Olivier and S.~Manuel,
	``Airy Functions And Applications To Physics (2nd Edition)''
	(World Scientific Publishing Company, 2010).
	

\bibitem{Barut:1974ch}
	A.~O.~Barut and D.~Villarroel,
	``Radiation Reaction and Mass Renormalization in Scalar and Tensor Fields and Linearized Gravitation,''
	J. \ Phys. \ A {\bf 8} (1975) 156.
	
\bibitem{Mironov:2006wi}
	A.~Mironov and A.~Morozov,
    ``Is Strong Gravitational Radiation predicted by TeV-Gravity?''
    JETP \ Lett. \ {\bf 85} (2007) 6
	[arXiv:0612074 [hep-ph]].
	
\bibitem{Bogush}
    I.~Bogush,
    (private communication).

 
\end{thebibliography}
\end{document}